\documentclass[aps,prd,twocolumn,showpacs, 
               amsmath,amssymb]{revtex4-1}

\usepackage{dsfont}
  \usepackage{bbold}
\usepackage{graphicx}

\def\be{\begin{equation}}
\def\ee{\end{equation}}
\def\beqn{\begin{eqnarray}}
\def\eeqn{\end{eqnarray}}
\def\no{\nonumber}
\def\ba{\begin{array}{c}}
\def\bat{\begin{array}{cc}}
\def\ea{\end{array}}
\def\bi{\begin{itemize}}
\def\ei{\end{itemize}}

\def\cL{{\cal L}}

\def\cO{{\cal O}}

\newcommand{\eqn}[1]{(\ref{#1})}
\newcommand{\bel}[1]{\be\label{#1}}

\begin{document}
\title{QCD exotics versus a Standard Model Higgs}

\author{Victor Ilisie} \author{Antonio Pich}
\affiliation{Departament de F\'{\i}sica Te\`orica, IFIC,
Universitat de Val\`encia -- CSIC,
Apt. Correus 22085, E-46071 Val\`encia, Spain}

\begin{abstract}
The present collider data put severe constraints on any type of new strongly-interacting
particle coupling to the Higgs boson. We analyze the phenomenological limits on exotic
quarks belonging to non-triplet $SU(3)_C$ representations and their implications on Higgs searches. The discovery
of the Standard Model Higgs, in the experimentally allowed mass range, would exclude the presence of exotic quarks coupling to it.
Thus, such QCD particles could only exist provided that their masses do not originate in the SM Higgs mechanism.
\end{abstract}

\date{22 February 2012}


\maketitle

\section{Exotic coloured fermions}

Exotic matter in higher representations of the $SU(3)_C$ colour group is an appealing
possibility which was already considered in the early times of QCD
\cite{MA:75,KA:76,WZ:77,NT:78,GG:79}.
In particular, the sextet representation has been extensively analyzed as a possible source of dynamical electroweak symmetry breaking \cite{MA:80,HP:82,KT:83,LPSZ:86,BWW:86,CLLR:86,WH:87,FMSWYI:91}.
It is well known that such exotic quarks modify very sizeably the running of the strong coupling and, therefore, their hypothetical existence is strongly constrained by the very successful experimental tests of asymptotic freedom \cite{alphas}.

Since not a single exotic QCD particle has been observed so far, their masses should
be heavy enough to avoid the present experimental constraints from direct searches.
However, even with very large masses,
if those exotic quarks get their masses through the Standard Model Higgs mechanism,
they would strongly enhance the production of Higgs bosons at LHC. The non-decoupling
character of the Higgs couplings, being proportional to the coupled-object mass,
implies sizeable effects from any strongly-interacting heavy mass scale generated by the Higgs mechanism. Therefore, the present collider limits on the production cross section
$\sigma(gg\to H)$ put a very severe constraint on the possible existence of such objects.

Let us consider an exotic spin-$\frac{1}{2}$ fermion $X_R$, with mass $M_X$, belonging
to the irreducible representation $\underline{R}\equiv (\lambda_1,\lambda_2)$ of $SU(3)_C$.
The dimension of the representation is given by
$d_R =\frac{1}{2}\, (\lambda_1 +1)(\lambda_2 +1)(\lambda_1 +\lambda_2 +2)$; the fundamental
$\underline{3}=(1,0)$ [$\underline{3}^*=(0,1)$] and adjoint
$\underline{8}=(1,1)$ representations have dimensions
$d_F=3$ and $d_A=8$, respectively.
The gluonic couplings of $X_R$ are fixed by the generators $t^a_R$ ($a=1,\cdots,d_A$), 
satisfying $[t^a_R,t^b_R]= i f^{abc}\, t^c_R$.
The quadratic Casimir operator,
\beqn\label{eq:Casimir}
\sum_{a=1}^{d_A}&&\!\!\!\! t^a_R\, t^a_R \; =\; C_R\,  \mathds{1}_{_{d_R}}\, ,
\no\\ 
C_R\; & =& \;\frac{1}{3}\, \left( \lambda_1^2+ \lambda_2^2 + \lambda_1\lambda_2 + 3 \lambda_1 + 3 \lambda_2\right)\, ,
\eeqn
determines the trace normalization factor for the representation $\underline{R}$:
\bel{eq:traceR}
\mathrm{Tr}\left(t^a_R\, t^b_R\right)\; =\; T_R\,\delta^{ab}\, ,
\qquad\qquad
T_R\; =\; \frac{C_R\, d_R}{d_A}\, .
\ee
This trace factor grows rapidly with increasing dimensions $d_R$, implying larger
contributions of the exotic object $X_R$ to the relevant QCD cross sections:
$T_F=\frac{1}{2}$, $T_6 = \frac{5}{2}$, $T_A=3$,
$T_{10} = \frac{15}{2}$, $T_{15} = 10$ \ldots,
where $\underline{6}=(2,0)$, $\underline{10}=(3,0)$, $\underline{15}=(2,1)$ \ldots

If kinematically allowed, charged exotic quarks would be copiously produced in
$e^+e^-$ annihilation. For a charged $X_R$ the ratio
$R_{e^+e^-}\equiv \sigma(e^+e^-\to\mathrm{hadrons})
/\sigma(e^+e^-\to\mu^+\mu^.)$
would rise dramatically at the production threshold $s= 4 M_X^2$  with an additive contribution $\Delta R_{e^+e^-} = d_R Q_X^2\delta_{_{\mathrm{QCD}}}$. A neutral
exotic $X_R^0$ would be pair-produced at $\cO(\alpha_s^2)$ through gluon emission,
i.e. $e^+e^-\to q\bar q g\to q\bar q X^0_R \bar X^0_R$. Independently of their electric charge, exotic quarks would imply large modifications of the hadronic cross sections at
$pp$ and $p\bar p$ colliders and a proliferation of new hadrons containing $X_R$
constituents (unless the $X_R$ lifetime is too small to hadronize). The absence of any exotic signal in the present data puts the lower
limit on the mass $M_X$ well above 100 or 200 GeV.

New fermions in higher QCD representations would contribute to the QCD $\beta$ function
\bel{eq:beta}
\mu\,\frac{d\alpha_s}{d\mu}\; =\; \alpha_s\,\beta (\alpha_s)\, ,
\qquad\quad
\beta (\alpha_s)\; =\; \sum_{n=1}\, \beta_n\,\left(\frac{\alpha_s}{\pi}\right)^n\, .
\ee
At the two loop level \cite{CA:74,JO:74},
\beqn
\beta_1 & = & -\frac{11}{6}\, C_A\, + \,\frac{2}{3}\,\sum_R  n_{_R}\, T_R\, ,
\no\\
\beta_2 & = & -\frac{17}{12}\, C_A^2\, + \,\frac{1}{6}\,\sum_R  n_{_R}\, T_R\,
(5\, C_A + 3\, C_R)\, ,
\eeqn
where $n_{_R}$ is the number of fermion flavours in the representation $\underline{R}$.
In the three-generation Standard Model
($n_F=6$) both $\beta_1$ and $\beta_2$ are negative. In order to flip the sign of
$\beta_1$ ($\beta_2$), $n_F> 16$ (8) triplet quarks would be needed. However,
the larger algebraic contribution of a higher colour representation implies a much faster lost of asymptotic freedom. Keeping $n_F=6$, the only possible additions preserving
$\beta_1<0$ are at most two sextet or one octet fermion representations; but even a single sextet flips already the sign of $\beta_2$. Since the running of $\alpha_s$ has been successfully tested with high precision (at the four loop level) from the $\tau$ mass scale
\cite{PI:11,BNP:92} up to energies above 200 GeV \cite{alphas}, exotic quarks in higher QCD representations are clearly excluded in this energy domain \cite{BE:04,CF:97,ALEPH:04,OPAL:11,JADE-OPAL:00}.

\begin{figure}[tb]\centering
\includegraphics[width=8cm]{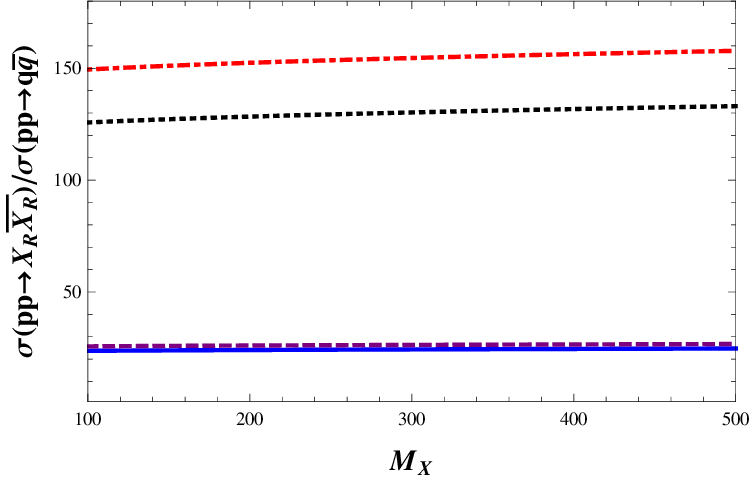}
\caption{\small Ratio $\sigma(pp\to X_R\bar X_R)/\sigma(pp\to q\bar q)$
at $\sqrt{s}= 7\:\mathrm{TeV}$, as a function of $M_X$.
The different curves correspond to the exotic fermion $X_R$
in the sextet (continuous, blue), octet (dashed, violet), decuplet (dotted, black)
and $\underline{15}$ (dash-dotted, red) representations.}
\label{fig:Xprod}
\end{figure}

Higher energy scales are presently being explored at the LHC, where the main production mechanism of exotic QCD fermions is $g g\to X_R\bar X_R$, with a subdominant contribution from
$q\bar q\to X_R\bar X_R$. The calculation of the corresponding partonic cross sections
is straightforward at tree level; we obtain
\beqn\label{eq:RprodG}
\sigma(gg\to X_R\bar X_R)& =& \frac{\pi\alpha_s^2}{16\, s}\, C_R\, d_R\;\,
{\cal G}\!\left(\frac{4 M_X^2}{s}\right) \, ,
\\
\label{eq:RprodQ}
\sigma(q\bar q\to X_R\bar X_R)& =& \frac{2\pi \alpha_s^2}{27\, s}\, C_R d_R
\left( 1 + \frac{2 M_X^2}{s}\right) \sqrt{1-\frac{4 M_X^2}{s}}\, ,
\no\eeqn
where
\beqn
{\cal G}(x)& =& \left[\left(1+x-\frac{x^2}{2}\right)  C_R +\frac{3}{4}\,  x^2\right]
 \ln{\left(\frac{1+\sqrt{1-x}}{1-\sqrt{1-x}}\right)}
\no\\
&- &\left[(1+x)\, C_R + 1 + \frac{5}{4}\,  x\right] \sqrt{1-x}\, ,
\eeqn
in agreement with Ref.~\cite{KRT:11}. Particularizing to the fundamental representation,
one gets the well-known results for quark-antiquark production \cite{ESW:03}.
The production of exotic fermions in higher representations is enhanced by
the global algebraic factor $\xi_R = C_R d_R/(C_F d_F)$
[$\xi_6 = 5$, $\xi_8 = 6$, $\xi_{10} = 15$, $\xi_{15} = 20$, \ldots],
which is further reinforced by another factor $C_R/C_F$ in the leading parts of the
2-gluon contribution. Figure~\ref{fig:Xprod} shows the ratio
$\sigma(pp\to X_R\bar X_R)/\sigma(pp\to q\bar q)$ at $\sqrt{s}= 7\;\mathrm{TeV}$,
as a function of $M_X$, for the representations with lower dimensions.
We have convoluted the partonic cross sections with
standard parton distribution functions and have assumed a common $K$ factor for all representations; i.e., we have taken the same QCD corrections as for triplet quark production.
This is a very conservative assumption because, given the larger algebraic factors,
gluonic corrections should be larger for higher colour representations. Thus, the curves
in Fig.~\ref{fig:Xprod} are actually lower bounds on the expected production ratios.
The enhancement factors are predicted to be larger than 10 for sextet and octet
fields and much higher values are obtained for higher-dimensional representations.

Once produced, the exotic $X_R$ particles should decay strongly generating an excess of
(multi) jet events. Fermionic objects in the triplet, sextet and  $\underline{15}$
representations could couple to a $qg$ ($\bar q g$) operator and are thus expected
to produce 2-jet events, while fermionic octets and decuplets have
$qqq$ ($\bar q\bar q\bar q$) quantum numbers and should be looked for
in 3-jet events \cite{KRT:11}.
The generic 2-jet searches performed at the LHC \cite{ATLAS:11a,CMS:11a}
have not found any evidence for new particle production, severely constraining
narrow resonances decaying into $qq$, $qg$ or $gg$ final states.
The lower limits on different types of strongly-interacting particles have been
pushed up beyond the 1 TeV scale; for instance the data excludes at 95\% CL
excited quarks with mass below 2.64 TeV or coloured octet scalars with mass
below 1.92 TeV.
Searches with 3 jets have been already performed by CMS \cite{CMS:11b} and
CDF \cite{CDF:11c}; no significant excess has been found, excluding gluino masses up to 280 GeV \footnote{Stronger bounds on gluino masses are obtained through searches for jet events with large missing energy or transverse momentum \cite{ATLAS:11d,CMS:11d,CDF:11d,D0:11d}.
The excluded region depends on the assumed supersymmetric model, reaching in some cases the 1 TeV scale.}.

A dedicated search for stable quarks in higher colour representations was
performed a long time ago by CDF \cite{CDF:89}. No such particles were found
in $26.2\;\mathrm{nb}^{-1}$ of data; at 95\% CL, the resulting lower
limits for $M_X$ were 98 (84) GeV for color sextets, 99
(86) GeV for octets, and 137 (121) GeV for decuplets,
assuming that $X_R$ carries charge one (either one or zero).
A recent CMS search for heavy stable charged particles produced
at LHC has put a lower limit of 808 GeV (95\% CL) on a stable gluino, under
the conservative hypothesis that any hadron containing this particle becomes neutral before
reaching the muon detectors (relaxing this hypothesis, the limit improves to 899 GeV)
\cite{CMS:11m}. Slightly weaker bounds have been set by ATLAS through a search for
slow-moving gluino-based R-hadrons \cite{ATLAS:11m}.

The present 95\% CL limits on fourth-generation quarks,
$m_{Q'} > 350\:\mathrm{GeV}$ \cite{ATLAS:11z},
$m_{b'} > 372\:\mathrm{GeV}$ \cite{CDF:11a} and
$m_{t'} > 404\:\mathrm{GeV}$ \cite{ATLAS:11w,CDF:11b,D0:11a}
assume the decays (with 100\% branching fraction) $Q'\to W q$,
$b'\to W t$ and $t'\to W b$, 
respectively.
While these direct limits are set on new triplet quarks,
the (absence of) experimental signature, $W+\mathrm{Jets}$, is also sensitive to other strongly-interacting exotic particles in weak $SU(2)_L$ representations, as we are going to consider next, provided they decay within the detector through
$X_R\to W X'_R\to W+\mathrm{Jets}$.

\section{Higgs production at LHC}

In the Standard Model, the Higgs mechanism is responsible for all particle masses.
If the mass of the exotic colour object $X_R$ is also generated through
its coupling to the Higgs boson, the Higgs properties are modified through quantum loops
involving the fermion $X_R$. Let us consider the consequences of a generic Higgs coupling
\bel{eq:Hcoup}
\cL_{H}\; =\; -
\frac{M_X}{v}\; H(x)\; \left[ \bar X_R(x)\, X_R(x)\right]\, ,
\ee
with $v =(\sqrt{2} G_F)^{-1/2} = 246$~GeV the Higgs vacuum expectation value.
The usual Standard Model mechanism for fermion masses requires $X_R$ to
be an electroweak doublet. More specifically, $X_R$ contains two fermion
fields, differing by one unit of electric charge, with their left-handed
chiralities forming a $SU(2)_L$ doublet while their right-handed chiralities are
singlets. We neglect their mass difference since the two fields should be
degenerated enough to satisfy the electroweak precision tests.
One should also implement the cancelation of the electroweak anomalies generated
by the new $SU(2)_L$ doublet;
we will assume for the moment that this is achieved through the addition of new
exotic leptons. We will comment later on the implications of arranging instead the anomaly cancelation
with additional coloured objects.
The anomaly constraints are discussed in the appendix for completeness.

%

Since $X_R$ couples strongly to gluons, the vertex in Eq.~\eqn{eq:Hcoup} generates a very sizeable contribution to the main Higgs production channel at LHC, through an intermediate
$X_R\bar X_R$ virtual pair: $gg\to X_R\bar X_R\to H$. The resulting amplitude can be easily obtained from the standard quark-loop result, accounting for the different colour factors:
%
%
\beqn\label{eq:Hgg}
\sigma (gg\to H)\, &=&\, \frac{M_H^2\alpha_s^2}{256\pi\, v^2}\;
\left|\, \sum_q T_F\, {\cal F}\!\left(\frac{4 m_q^2}{M_H^2}\right)
\right.\no\\ &&\left.\mbox{}
+ 2\, T_R \, {\cal F}\!\left(\frac{4 M_X^2}{M_H^2}\right)\,\right|^2\;
\delta(s-M_H^2)\, ,
\eeqn
where
\beqn
{\cal F}(x)\, & =&\, \frac{x}{2}\,\left[ 4 + (x-1) f(x)\right]\, ,
\no\\
f(x)\, &=&\, \left\{ \!\bat
-4\, \arcsin^2{(1/\sqrt{x})}\, , &\quad x\ge 1
\\[3pt]
\left[ \ln{\left(\frac{1+\sqrt{1-x}}{1-\sqrt{1-x}}\right)}-i \pi\right]^2 ,
&\quad x < 1\ea\right.\, .
\eeqn
The first term in \eqn{eq:Hgg} is the usual triplet-quark contribution;
it is completely dominated by the top loop because
the function ${\cal F}(x)$ vanishes in the massless limit ($x\to 0$).
The second term stands for the
additional contribution from the exotic coloured fermion multiplet $X_R$.
Given the experimental constraints on $M_X$ discussed before, $M_H^2 < 4 M_X^2$ in the interesting kinematical regime and the corresponding loop function does not have any
absorptive part. Moreover, the numerical result is not sensitive to the exact value of $M_X$ because
${\cal F}(x)$ is a very smooth function for $x\ge 1$, decreasing gently from ${\cal F}(1)=2$
to ${\cal F}(\infty)=4/3$.

\begin{figure}[tb]\centering
\includegraphics[width=8cm]{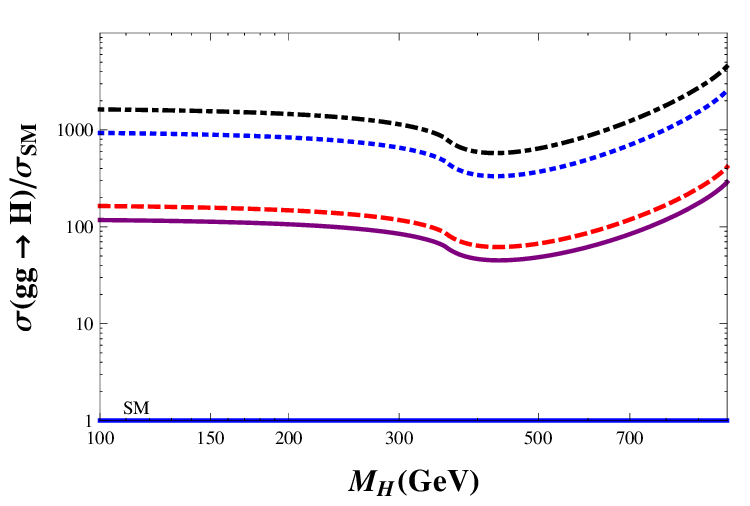}
\caption{\small Ratio $\sigma(gg\to H)/\sigma_{\mathrm{SM}}$
at $\sqrt{s}= 7\:\mathrm{TeV}$ and $M_X = 500$ GeV, as a function of $M_X$.
The different curves correspond to an exotic fermion multiplet $X_R$
in the sextet (continuous), octet (dashed), decuplet (dotted)
and $\underline{15}$ (dash-dotted) representations.}
\label{fig:Hprod}
\end{figure}

Owing to the relative colour enhancement factor $T_R/T_F$,
the $X_R$ contribution generates
a large increase of the Higgs production cross section. The ratio
$\sigma(gg\to H)/\sigma_{\mathrm{SM}}$ for different colour representations
is shown in Fig.~\ref{fig:Hprod}, as a function of $M_H$, taking $\sqrt{s}= 7\:\mathrm{TeV}$
and $M_X = 500$ GeV. The normalization $\sigma_{\mathrm{SM}}\equiv\sigma(gg\to H)_{\mathrm{SM}}$ is the Standard Model
cross section with three quark families. Again, we have assumed the same QCD corrections
as for triplet quarks, which underestimates the actual cross section.
Very large enhancement factors are obtained for all non-triplet representations. In the sextet and octet cases,
the Higgs production cross section is larger than the SM one by a factor
between 40 or 300, depending on $M_H$. The enhancement
surpasses the three orders of magnitude for the $\underline{15}$ and higher colour
representations.

\section{Higgs search}

\begin{figure}[tb]
\includegraphics[width=8cm]{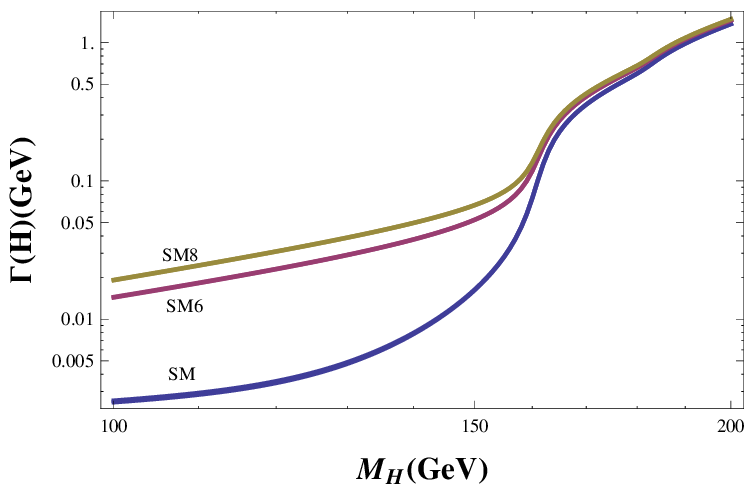}
\caption{\small Higgs total decay width in the 3-generation Standard Model (SM), and with the addition of colour sextet (SM6) or octet (SM8) multiplets.}
\label{fig:Hwidth}
\end{figure}

\begin{figure}[tb]
\includegraphics[width=8cm]{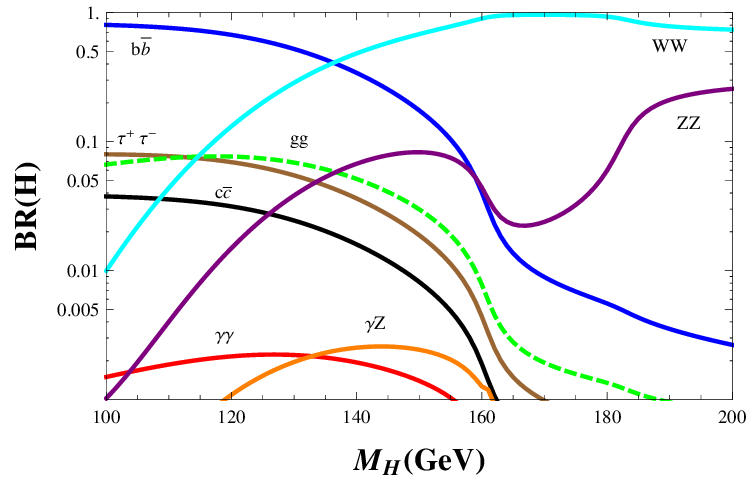}
\caption{\small Higgs decay branching ratios in the 3-generation Standard Model.}
\label{fig:Hbr-SM}
\end{figure}

\begin{figure}[tb]
\includegraphics[width=8cm]{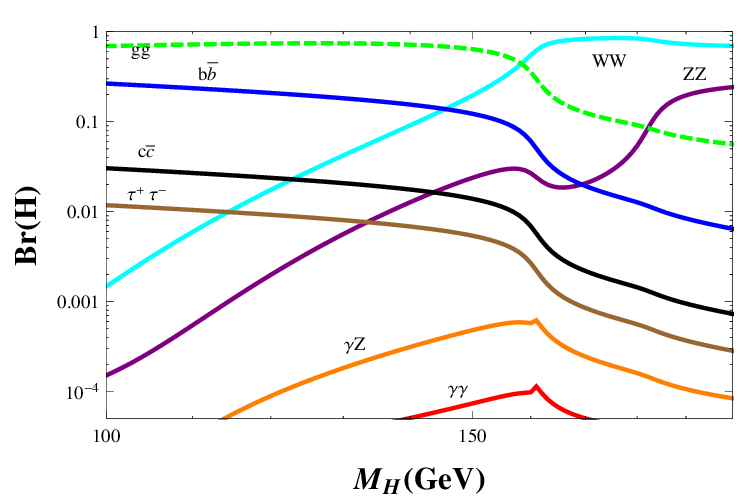}
\caption{\small Higgs branching ratios with the addition of a colour sextet multiplet.}
\label{fig:Hbr-SM6}
\end{figure}

\begin{figure}[tb]
\includegraphics[width=8cm]{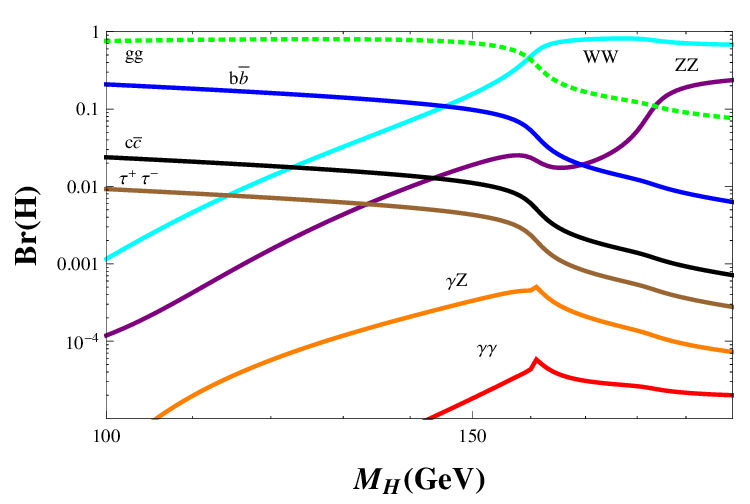}
\caption{\small Higgs branching ratios with the addition of a colour octet multiplet.}
\label{fig:Hbr-SM8}
\end{figure}

Since the decay $H\to X_R \bar X_R$ is not kinematically allowed for $M_H < 2 M_X$,
a heavy Higgs would decay into $WW$, $ZZ$ and $t\bar t$ with approximately the
same branching fractions as in the absence of the fermion $X_R$. The Standard Model
Higgs has already been experimentally excluded for Higgs masses
between $2 M_W$ and 600 (525) GeV, at 95\% CL (99\% CL)
\cite{ATLAS:11e,CMS:11e}.
The existence of an additional coloured fermion would only make the exclusion much stronger.
More care has to be taken below the $WW$ threshold, because the same enhancement
present in the Higgs production cross section also appears in the $H\to gg$
decay width, modifying all branching ratios. Figure~\ref{fig:Hwidth} shows
the total Higgs decay width $\Gamma_H$, as a function of $M_H$, for the Standard Model with
three families of triplet quarks, and with the addition of one
(electroweak doublet) colour sextet
or octet multiplet. The exotic contributions are small for $M_H>2 M_W$, but at lower Higgs masses they generate a big enhancement of $\Gamma_H$.
Figures~\ref{fig:Hbr-SM}, \ref{fig:Hbr-SM6} and \ref{fig:Hbr-SM8}
plot the corresponding branching ratios in the different channels.

\begin{figure}[t]\centering
\includegraphics[width=8cm]{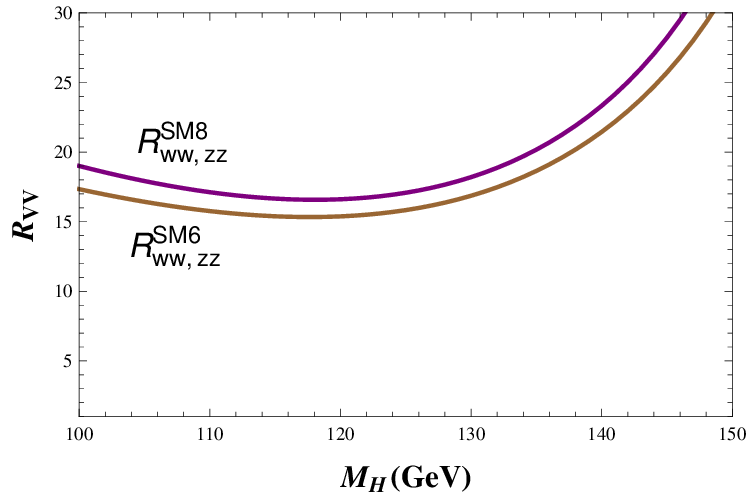}
\caption{\small $R_{WW,ZZ}$ at $\sqrt{s} = 7\;\mathrm{TeV}$, as a function of $M_H$, with the addition of colour sextet (SM6) or octet (SM8) multiplets.}
\label{fig:Rratio}
\end{figure}

The strong enhancement of the two-gluon decay channel at low Higgs masses, affects in a
very sizeable way the suppressed (one loop) $2\gamma$ and $\gamma Z$ decay modes, making
them insignificant.
However, in the $WW$ and $ZZ$ modes the branching fraction suppression
cannot compensate the large enhancement of the production rate. In order to compare
with the LHC experimental data, the relevant ratio is
\bel{eq:Rratio}
R_{VV}\; =\; \frac{\sigma(pp\to H) \;\mathrm{Br}(H\to VV)}{\sigma(pp\to H)_{\mathrm{SM}} \;\mathrm{Br}(H\to VV)_{\mathrm{SM}}}\, ,
\ee
where SM refers again to the Standard Model with three quark families and $V=W,Z$.
This is plotted in Fig.~\ref{fig:Rratio}, for sextet and octet colour representations, showing that, at $\sqrt{s} = 7\;\mathrm{TeV}$,
$R_{VV} > 15$ in the relevant range of Higgs masses.
Much larger values of $R_{VV}$ would be obtained with higher-dimensional representations
or additional coloured fermion multiplets.
Therefore,
the present ATLAS \cite{ATLAS:11e} and CMS \cite{CMS:11e} searches in the $WW$ and $ZZ$ channels,
already exclude a Standard Model Higgs boson coupled to exotic colour multiplets, in the whole range between 110 and 600 GeV.

The combined CDF and D0 data \cite{TevatronHiggs:11} exclude Higgs masses between 100 and 108 GeV (95\% CL), within the three-generation Standard Model.
Although $gg\to H$ accounts for 76\% of the
Higgs production cross section in this mass region, the Tevatron constraints are mainly
extracted from $q\bar q\to W H / ZH$, with a small contribution from
$q\bar q\to q'\bar q' H$. These production mechanisms are not enhanced
by the exotic colour-multiplet contributions. In this mass range the main Higgs signature
is $H\to b\bar b$; therefore,
the Tevatron information translates into 95\% CL upper bounds for
${\cal R}_{b\bar b}\equiv\mathrm{Br}(H\to b\bar b)/\mathrm{Br}(H\to b\bar b)_{\mathrm{SM}}$ ranging from 0.45 at 100 GeV
to 1.1 at 110 GeV \cite{TevatronHiggs:11}. The addition of a sextet (octet) multiplet
implies ${\cal R}_{b\bar b}$ values ranging from 0.33 (0.26) at 100 GeV to
0.31 (0.24) at 110 GeV, which are slightly below the present Tevatron bounds.
A mild improvement of the Tevatron constraints could exclude sextet ot octet contributions for $M_H$ between 100 and 110 GeV.

\begin{figure}[tb]\centering
\includegraphics[width=7.8cm]{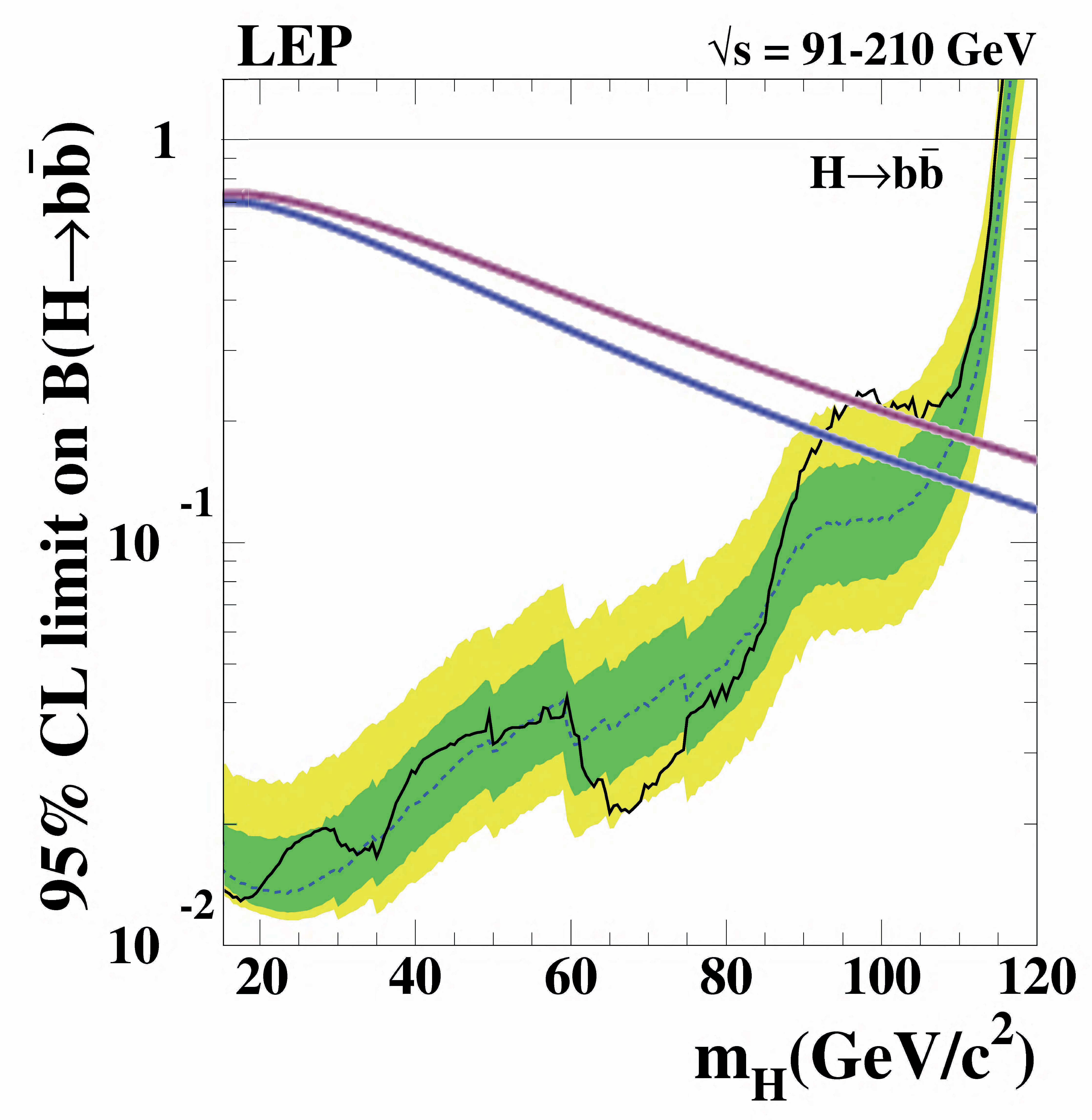}
\caption{\small The LEP exclusion limits on $\mathrm{Br}(H\to b\bar b)$ \cite{LEP:03},
as a function of $M_H$, are  compared with the expected signals in the presence of one exotic
(electroweak doublet) sextet (top red curve) or octet (bottom blue curve) multiplets.}
\label{fig:Rbb}
\end{figure}

The LEP exclusion limit below 114.5 GeV \cite{LEP:03} needs also to be re-analyzed in view of the strong
enhancement of $\mathrm{Br}(H\to gg)$. While the production mechanism $e^+e^-\to Z^*\to Z H$ remains unchanged in the presence of exotic quarks,
there is a large suppression of the Higgs branching fractions into $b\bar b$ and $\tau^+\tau^-$
and, therefore, of the sought experimental signal.
OPAL performed a generic search for neutral scalars decaying into an arbitrary combination of hadrons, leptons, photons and
invisible particles, covering as well the possibility of a stable scalar \cite{OPAL:03}. Thus, the OPAL bound,
$M_H > 81\;\mathrm{GeV}$ (95\% CL) \cite{OPAL:03}, remains valid in the presence of exotic colour multiplets.
For larger masses, the combined LEP analysis relies in the $H\to b\bar b$ decay mode.
Figure~\ref{fig:Rbb} compares the LEP bounds on $\mathrm{Br}(H\to b\bar b)$ \cite{LEP:03},
with the expected values with one (electroweak doublet) sextet (top red curve) or octet (bottom blue curve) multiplet.
Higgs masses below 96 (92) GeV are then excluded in the sextet (octet) case.

The triplet case of a fourth quark generation has been already discussed before
\cite{KPST:07,BCIS:10,CDF-D0:10,CMS:11x,ATLAS:11x,RV:10,BFKKS:03,KS:11,ABF:10,ABFHL:11,RZ:11,GU:11,HWG:11}.
The enhancement of $\sigma(gg\to H)$ is milder, about a factor of 9, but enough to
exclude Higgs masses above 110 GeV from the LHC constraints on $R_{VV}$. The corresponding
weaker enhancement of $\mathrm{Br}(H\to gg)$ implies a much smaller suppression of
the remaining channels; in particular, for Higgs masses smaller than 110 GeV, the $b\bar b$ branching fraction is predicted to be above the LEP bound in Fig.~\ref{fig:Rbb}.
Therefore, in the presence of an additional (electroweak doublet) colour quark triplet, the Higgs boson is excluded in the whole mass range up to 600 GeV.

Note, however, that additional exotic multiplets
or higher colour representations would imply a larger suppression of $\mathrm{Br}(H\to b\bar b)$, weakening the LEP and Tevatron constraints. That would be the case, for instance, if the anomaly matching condition is fulfilled with (at least two) coloured exotic multiplets, instead of leptons.
Thus, in the region of Higgs masses between 81 and 110 GeV the constraints are sensitive to the assumed exotic spectrum. This is not the case for lower or higher values of $M_H$;
Higgs masses between 110 and 600 GeV,
or smaller than 81 GeV, are excluded in the presence of any exotic colour multiplets coupled to the Higgs boson.

\section{Discussion}

Present LHC data imply that a Standard Model Higgs cannot exist in the presence of new
coloured fermions coupled to it, in exotic QCD representations,
except for a small $M_H$ region between 92 (81 with several exotic multiplets)
and 110 GeV which could be soon excluded.
Exotic quarks in higher-dimension colour representations generate a very large enhancement
of $\sigma(gg\to H)$, in contradiction with the available experimental bounds.
Strong limits have been already put before in the case of a fourth quark generation, where the enhancement of the Higgs production cross section is milder \cite{CDF-D0:10,CMS:11x,ATLAS:11x}.

One could certainly try to evade the experimental constraints, enlarging the
Standard Model in appropriate ways to compensate the enhancement from exotic quarks.
For instance, introducing additional coloured scalars with couplings to the Higgs
adjusted to suppress the $gg \to H$ amplitude \cite{DKM:11,HV:11,MW:06,DJ:98,BFH:11}.
Another possibility is ``hiding'' the Higgs; i.e., opening
new decay channels into invisible modes without strong interactions
\cite{BFKKS:03,KS:11,SS:82,HHTT:10,RS:11,MA:12,CDGW:08,BLPRZZ:10,EJRS:11},
in order to suppress the visible branching fractions.
While well-motivated arguments, such as dark matter, exist to do it, we feel that this
hides the main reason behind such strong exclusion: the intrinsic non-decoupling of the
Yukawa vertex \eqn{eq:Hcoup} makes the Higgs boson sensitive to arbitrary high mass scales.

The Higgs vacuum expectation value is linked to the electroweak scale, i.e., to the
gauge boson masses $M_W$ and $M_Z$. In the Standard Model this scale is also used to
generate all fermion masses through the Yukawa couplings. The known pattern of lepton and quark masses, with very different mass scales, implies a large variety of Yukawa couplings
with magnitudes ranging from $m_\nu/v\sim 10^{-13}$ to  $m_t/v\sim 0.7$.
This wide range of couplings/scales is not yet understood.
Introducing additional fermions with even higher masses, would bring
much larger Yukawa couplings inducing a non-perturbative dynamical regime in the electroweak
sector. In fact, the Higgs production and decay amplitudes used in our analysis are
subject to potentially large electroweak corrections \cite{HWG:11}.

If a light neutral scalar boson is finally discovered, one should study very carefully
its properties in order to clarify the true pattern of electroweak symmetry breaking.
The Standard Model is certainly a very plausible possibility, but
heavier mass scales should not couple to the Higgs boson, i.e., they should
have a different origin. Multi-Higgs models offer a much more flexible framework to
accommodate future data, but soon or later they would also face the characteristic non-decoupling of the Higgs mechanism in (parts of) their extended Yukawa couplings.
A perhaps more interesting possibility is that fermion masses could be generated
through a mechanism different than the one responsible for the gauge boson masses.
Another alternative, of course, is that the Higgs boson does not exist (dynamical
symmetry breaking) or it is a composite object with rather different properties.
The forthcoming LHC data should soon show us the option chosen by Nature to break
the electroweak symmetry and hopefully provide some hints on the dynamics behind
the observed pattern of fermion masses and mixings.

\section*{Acknowledgements}
This work has been supported in part by MICINN, Spain
[Grants FPA2007-60323, FPA2011-23778 and Consolider-Ingenio 2010 Program CSD2007-00042 (CPAN)] and by Generalitat Valenciana under Grant No. Prometeo/2008/069.

\section*{Appendix}

The cancelation of the triangular gauge anomalies requires \cite{PI:12}
\bel{eq:anomaly}
\mathrm{Tr}\left( \left\{T^a,T^b\right\} T^c\right)_L
- \mathrm{Tr}\left( \left\{T^a,T^b\right\} T^c\right)_R \; =\; 0\, ,
\ee
where $T^a$ are the Standard Model group generators and
the traces sum over all possible left- and right-handed fermions.
Owing to the algebraic properties of the $SU(2)$ and $SU(3)$ generators,
the only non-trivial anomalies involve one or three $U(1)_Y$ bosons, giving
conditions on traces of $Y$ and $Y^3$, respectively,
where the hypercharge is related to the electric charge through $Y=Q-T^3$.
These relations imply that the sum of all fermion
electric charges should be zero:
\bel{eq:anomaly3}
\sum_f Q_f\; =\; \mathrm{Tr}\left(Y\right)_L \; =\; \mathrm{Tr}\left(Y\right)_R\; =\; 0\, .
\ee

Let us consider $N$ $SU(2)_L$ fermion doublets $\psi_i$ with
$Y(\psi_{i,L}) = y_i$, and their corresponding right-handed singlets with $Y(\psi_{i,R}) = Q_i = y_i+\frac{1}{2}$ and $Y(\psi'_{i,R}) = Q'_i = y_i-\frac{1}{2}$. In order to cancel the Standard Model gauge anomalies, one needs to satisfy
\bel{eq:anomaly4}
2\, \sum_i^N\, d_i\, y_i\; =\; \sum_i^N\, d_i\, (2 Q_i -1) \; =\; 0\, ,
\ee
where $d_i$ denotes the multiplicity of the $SU(3)_C$ representation of $\psi_i$.
The number of left-handed fermion doublets, $\sum_i^N d_i$, should be even in order to avoid a
global (non-perturbative) $SU(2)$ chiral gauge anomaly \cite{WI:92}.
The normal Standard Model generations fulfill these conditions with one quark
($d_q = 3$, $y_q = \frac{1}{6}$) and one lepton ($d_l = 1$, $y_l = -\frac{1}{2}$)
multiplets.

Thus, there are many possible ways of adding exotic coloured fermions to the 3-generation Standard Model, while preserving the anomaly cancelation conditions.
A single exotic representation with even dimension and $y=0$ ($Q=\frac{1}{2}$, $Q' = -\frac{1}{2}$)
would of course be anomaly free, but it would be stable (it cannot decay into ordinary quarks and gluons). The simplest solution to the anomaly constraint involves two exotic multiplets with the
same $SU(3)_C$ multiplicity and opposite hypercharge.

The most general solution with two additional multiplets of different dimensionalities
is $y_2= - y_1 d_1 /d_2$, with $d_1+d_2$ even. For odd-dimensional exotic representations
($d_1=15, 27 \ldots$), it is then possible to cancel the anomaly with a new lepton multiplet
of hypercharge $y_2 = - y_1 d_1$.
Two lepton multiplets with $y_2 + y_3= - y_1 d_1$ would be needed to cancel the anomaly
of an exotic representation with even multiplicity ($d_1 = 6, 8, 10\ldots$).
For any exotic colour representation of dimension $d$ and hypercharge $y$, the anomaly could of course be canceled with $d$ lepton multiplets of hypercharge $y_l=-y$.

The figures shown in the paper refer to the simplest case of a single (electroweak doublet) exotic quark multiplet, with the anomaly canceled by exotic lepton
multiplets. If one considers instead models where the anomaly is canceled through additional coloured fermions,
the LHC constraints become much stronger in the whole mass range analyzed.
For instance two exotic quark multiplets with the same $SU(3)_C$ multiplicity and opposite hypercharge, would increase the ratio $R_{VV}$ (Fig. 7) by a factor close to two.
Therefore the range of Higgs masses between 110 and 600 GeV is completely excluded in any exotic model.
However, since additional coloured fermions imply a suppression of $\mathrm{Br}(H\to b\bar b)$, weakening the LEP and Tevatron constraints, an open window of allowed Higgs masses between 81 and 110 GeV remains in this type of models.


\begin{thebibliography}{99}

\bibitem{MA:75} E. Ma, {\it Phys. Lett.} {\bf 58B} (1975) 442.

\bibitem{KA:76} G. Karl, {\it Phys. Rev.} {\bf D14} (1976) 2374.

\bibitem{WZ:77} F. Wilczek and A. Zee, {\it Phys. Rev.} {\bf D16} (1977) 860.

\bibitem{NT:78} Y. Ng and S.-H. Tye, {\it Phys. Rev. Lett.} {\bf 41} (1978) 6.

\bibitem{GG:79} H. Georgi and S. Glashow, {\it Nucl. Phys.} {\bf B159} (1979) 29.

\bibitem{MA:80} W.J. Marciano, {\it Phys. Rev.} {\bf D21} (1980) 2425.

\bibitem{HP:82} B. Holdom and M.E. Peskin, {\it Nucl. Phys.} {\bf B208} (1982) 397.

\bibitem{KT:83} K. Konishi and R. Tripiccione, {\it Phys. Lett.} {\bf B121} (1983)
403.

\bibitem{LPSZ:86} D. L\"ust et al.,  
{\it Nucl. Phys.} {\bf B268} (1986) 49.

\bibitem{BWW:86} E. Braaten, A.R. White and C.R. Willcox,
{\it Intern. J. Mod. Phys.} {\bf A1} (1986) 693.

\bibitem{CLLR:86} T.E. Clark et al.,  
{\it Phys. Lett.} {\bf B177} (1986) 413.

\bibitem{WH:87} A.R. White, {\it Mod. Phys. Lett.} {\bf A2} (1987) 945.

\bibitem{FMSWYI:91} K. Fukazawa et al., {\it Prog. Theor. Phys.} {\bf 85} (1991) 111.

\bibitem{alphas} S. Bethke et al., arXiv:1110.0016 [hep-ph].

\bibitem{CA:74} W.E. Caswell, {\it Phys. Rev. Lett.} {\bf 33} (1974) 244.

\bibitem{JO:74} D.R.T. Jones, {\it Nucl. Phys. Rev.} {\bf B75} (1974) 531.


\bibitem{PI:11} A. Pich, arXiv:1107.1123 [hep-ph].

\bibitem{BNP:92} E. Braaten, S. Narison and A. Pich, {\it Nucl. Phys.} {\bf B373} (1992) 581.

\bibitem{BE:04}	S. Bethke, {\it Phys. Rept.} {\bf 403-404} (2004) 203;
{\it Eur. Phys. J.} {\bf C64} (2009) 689.

\bibitem{CF:97} F. Csikor and Z. Fodor,
{\it Phys. Rev. Lett.} {\bf 78} (1997) 4335.

\bibitem{ALEPH:04} ALEPH Collaboration,
{\it Eur. Phys. J.} {\bf C35} (2004) 457;
%
{\bf C27} (2003) 1;
{\it Z. Phys.} {\bf C76} (1997) 1;
%
{\it Phys. Rept.} {\bf 294} (1998).

\bibitem{OPAL:11} OPAL Collaboration,
{\it Eur. Phys. J.} {\bf C71} (2011) 1733;  
{\bf C20} (2001) 601;   
{\bf C16} (2000) 185.   

\bibitem{JADE-OPAL:00} JADE and OPAL Collaborations,	
{\it Eur. Phys. J.} {\bf C17} (2000) 19.

\bibitem{KRT:11} J. Kumar, A. Rajaraman and B. Thomas,
{\it Phys. Rev.} {\bf D84} (2011) 115005.

\bibitem{ESW:03}
R.K. Ellis, W.J. Stirling and B.R. Webber, {\it QCD and Collider Physics}, Cambridge Monographs in Particle Physics, Nuclear Physics and Cosmology
(Cambridge University Press, 2003).



\bibitem{ATLAS:11a} ATLAS Collaboration,
{\it New J. Phys.} {\bf 13} (2011) 053044;
%
{\it Phys. Lett.} {\bf B} 708 (2012) 37.

\bibitem{CMS:11a} CMS Collaboration,	
{\it Phys. Lett.} {\bf B704} (2011) 123;
%
{\it Phys. Rev. Lett.} {\bf 105} (2010) 211801.

\bibitem{CMS:11b} CMS Collaboration,
{\it Phys. Rev. Lett.} {\bf 107} (2011) 101801.

\bibitem{CDF:11c} CDF Collaboration,
{\it Phys. Rev. Lett.} {\bf 107} (2011) 042001.

\bibitem{ATLAS:11d} ATLAS Collaboration,
{\it Phys. Lett.} {\bf B710} (2012) 67;  
{\bf B701} (2011) 186, 
%
398;
%
{\it Phys. Rev. Lett.} {\bf 106} (2011) 131802; 
%
%
{\it Eur. Phys. J.} {\bf C71} (2011)  1682.  

\bibitem{CMS:11d} CMS Collaboration,
{\it Phys. Lett.} {\bf B698} (2011) 196;  
%
{\it JHEP} {\bf 1108} (2011) 155;    
%
{\it Phys. Rev.} {\bf D85} (2012) 012004;  
%
{\it Phys. Rev. Lett.} {\bf 107} (2011) 221804;  
%
{\bf 106} (2011) 211802;
%
{\it JHEP} {\bf 07} (2011) 113.

\bibitem{CDF:11d} CDF Collaboration,
%
{\it Phys. Rev. Lett.} {\bf 101} (2008) 251801;
%
{\bf 102} (2009) 121801. 	

\bibitem{D0:11d} D0 Collaboration,
%
{\it Phys. Lett.} {\bf B660} (2008) 449;
%
{\bf B680} (2009) 34. 	

\bibitem{CDF:89} CDF Collaboration, {\it Phys. Rev. Lett.} {\bf 63} (1989) 1447.

\bibitem{CMS:11m} CMS Collaboration,
CMS PAS EXO-11-022 (2011);
%
{\it JHEP} {\bf 1103} (2011) 024;
%
{\it Phys. Rev. Lett.} {\bf 106} (2011) 011801.

\bibitem{ATLAS:11m} ATLAS Collaboration,
{\it Phys. Lett.} {\bf B701} (2011) 1;
%
{\bf B703} (2011) 428;
%
{\it Eur. Phys. J.} {\bf C72} (2012) 1965.


\bibitem{ATLAS:11z} ATLAS Collaboration,
arXiv:1202.3389 [hep-ex].

\bibitem{CDF:11a} CDF Collaboration,
{\it Phys. Rev. Lett.} {\bf 106} (2011) 141803.

\bibitem{ATLAS:11w} ATLAS Collaboration,
{\it Phys. Rev. Lett.} {\bf 108} (2012) 261802. 

\bibitem{CDF:11b} CDF Collaboration,
{\it Phys. Rev. Lett.} {\bf 107} (2011) 261801.    

\bibitem{D0:11a} D0 Collaboration,
{\it Phys. Rev. Lett.} {\bf 107} (2011) 082001.  




\bibitem{ATLAS:11e} ATLAS Collaboration,
{\it Phys. Lett.} {\bf B710} (2012) 49,  
383;    
{\it Phys. Rev. Lett.} {\bf 108} (2012) 111802,  
111803. 

\bibitem{CMS:11e} CMS Collaboration,
{\it Phys. Lett.} {\bf B710} (2012) 26,   
91, 
284, 
403;  
{\bf B713} (2012) 68; 
%
{\it JHEP} {\bf 1203} (2012) 040,   
081; 
{\bf 1204} (2012) 036;    
{\it Phys. Rev. Lett.} {\bf 108} (2012) 111804.    

\bibitem{TevatronHiggs:11}
TEVNPH (Tevatron New Phenomena and Higgs Working Group) and CDF and D0 Collaborations,
arXiv:1107.5518 [hep-ex].

\bibitem{LEP:03} ALEPH, DELPHI, L3 and OPAL Collaborations,
The LEP Working Group for Higgs Boson Searches,
{\it Phys. Lett.} {\bf B565} (2003) 61.

\bibitem{OPAL:03} OPAL Collaboration, {\it Eur. Phys. J.} {\bf C27} (2003) 311.



\bibitem{KPST:07}
G.D. Kribs, T. Plehn, M. Spannowsky and T.M.P. Tait,
{\it Phys. Rev.} {\bf D76} (2007) 075016.

\bibitem{BCIS:10}
N. Becerici Schmidt, S.A. \c{C}etin, S. I\c{s}tin and S. Sultansoy,
{\it Eur. Phys. J.} {\bf C66} (2010) 119.  

\bibitem{CDF-D0:10} CDF and D0 Collaborations,
{\it Phys. Rev.} {\bf D82} (2010) 011102.  

\bibitem{CMS:11x} CMS Collaboration,
{\it Phys. Lett.} {\bf B699} (2011) 25; 
CMS-PAS-HIG-11-011.

\bibitem{ATLAS:11x} ATLAS Collaboration,
{\it Eur. Phys. J.} {\bf C71} (2011) 1728.    

\bibitem{RV:10} A.N. Rozanov and M.I. Vysotsky,
{\it Phys. Lett.} {\bf B700} (2011) 313.

\bibitem{BFKKS:03}
K. Belotsky et al., 
{\it Phys. Rev.} {\bf D68} (2003) 054027.   

\bibitem{KS:11} W.-Y. Keung and P. Schwaller,
{\it JHEP} {\bf 1106} (2011) 054.   

\bibitem{ABF:10}	
C. Anastasiou, R. Boughezal and E. Furlan, {\it JHEP} {\bf 1006} (2010) 101.

\bibitem{ABFHL:11}
C. Anastasiou et al.,  
{\it Phys. Lett.} {\bf B702} (2011) 224.    

\bibitem{RZ:11}	
X. Ruan and Z. Zhang, arXiv:1105.1634 [hep-ph].

\bibitem{GU:11}	
J.F. Gunion, arXiv:1105.3965 [hep-ph].

\bibitem{HWG:11}	
A. Denner et al., {\it Eur. Phys. J.} {\bf C72} (2012) 1992.



\bibitem{DKM:11}	
B.A. Dobrescu, G.D. Kribs and A. Martin,    
{\it Phys. Rev.} {\bf D85} (2012) 074031.

\bibitem{HV:11} X.-G. He and G. Valencia,
{\it Phys. Lett.} {\bf B707} (2012) 381.  

\bibitem{MW:06} A.V. Manohar and M.B. Wise,
{\it Phys. Rev.} {\bf D74} (2006) 035009.   

\bibitem{DJ:98}	
A. Djouadi, {\it Phys. Lett.} {\bf B435} (1998).

\bibitem{BFH:11}		
Y. Bai, J.J. Fan, J.L. Hewett, arXiv:1112.1964 [hep-ph].


\bibitem{SS:82}	
R.E. Shrock and M. Suzuki, {\it Phys. Lett.} {\bf B110} (1982) 250.

\bibitem{HHTT:10}
X.-G. He, S.-Y. Ho, J. Tandean and H.-C. Tsai, {\it Phys. Rev.} {\bf D82} (2010) 035016.

\bibitem{RS:11}	
M. Raidal and A. Strumia, {\it Phys. Rev.} {\bf D84} (2011) 077701.

\bibitem{MA:12}	E. Ma,
{\it Phys. Lett.} {\bf B706} (2012) 350;    
{\it Int. J. Mod. Phys.} {\bf A27} (2012) 1250059.

\bibitem{CDGW:08}		
S. Chang, R. Dermisek, J.F. Gunion and N. Weiner,
{\it Ann. Rev. Nucl. Part. Sci.} {\bf 58} (2008) 75.

\bibitem{BLPRZZ:10}	
S. Bock et al.,   
{\it Phys. Lett.} {\bf B694} (2010).

\bibitem{EJRS:11} C. Englert et al.,  
{\it Phys. Rev.} {\bf D85} (2012) 035008.  

\bibitem{PI:12} A. Pich, ``The Standard Model of Electroweak Interactions'',
Proc. 2010 European School of High-Energy Physics, CERN-2012-001, p.~1,
arXiv:1201.0537 [hep-ph].

\bibitem{WI:92} E. Witten, {\it Phys. Lett.} {\bf B117} (1982) 324.

\end{thebibliography}
\end{document}